\documentclass[aps,pre,floatfix,superscriptaddress,amsmath,showpacs,twocolumn]{revtex4}
\usepackage{graphicx}
\newcommand {\be}{\begin{equation}}
\newcommand {\ee}{\end{equation}}

\begin{document}

\title{Resonant activation in bistable semiconductor lasers}

\author{Stefano Lepri}
\email{stefano.lepri@isc.cnr.it}
\author{Giovanni Giacomelli}
\affiliation{Istituto dei Sistemi Complessi, Consiglio Nazionale
delle Ricerche, via Madonna del Piano 10, I-50019 Sesto Fiorentino, Italy}

\date{\today}

\begin{abstract}
We theoretically investigate the possibility of observing  resonant
activation in the hopping dynamics of two-mode semiconductor lasers. 
We present a series of simulations of a rate-equations model under
random and periodic modulation of the bias current. In both cases, 
for an optimal choice of the modulation time-scale, the hopping times 
between the stable lasing modes attain a minimum. The simulation
data are understood by means of an effective one-dimensional 
Langevin equation with multiplicative fluctuations. Our
conclusions apply to both Edge Emitting and Vertical Cavity Lasers, thus
opening the way to several experimental tests in such optical systems. 
\end{abstract}
\maketitle

\section{Introduction}

It is currently established that stochastic fluctuations may have a constructive
role in enhancing the response of nonlinear systems to an external coherent 
stimulus. Relevant examples are the enhancement of the decay time from a
metastable state (noise--enhanced stability) \cite{Graham,Mantegna}, the
synchronization with a weak periodic input signal (stochastic resonance)
\cite{sto_res} or the regularizaton of the response  at an optimal noise
intensity (coherence resonance) \cite{coh_res}. 

Another instance is the phenomenon of resonant activation that was 
discovered by Doering and Gadoua \cite{doering}. They showed that 
the escape of an overdamped Brownian particle over a fluctuating barrier
can be enhanced by suitably choosing the correlation time of  barrier
fluctuations themselves. In other words, the escape time from the
potential well attains a minimum for an optimal choice of such
correlation time. Since its discovery, the phenomenon received a
considerable attention from theorists (see e.g.
Refs.~\cite{ba93,h94,h95,r95,iw96}). Detailed studies by means of analog
simulations have also been reported for both Gaussian and dichotomous
fluctuations \cite{m96}. More recently, the phenomenon has been shown
to occur also for the case in which the barrier oscillates 
periodically \cite{salerno,chaos}.

To our knowledge, experimental evidences of resonant activation were
only given for a bistable electronic circuit \cite{maspa} and, very
recently, for a colloidal particle subject to a
periodically--modulated optical potential \cite{schmitt}. It is
therefore important to look for other setups where the effect could be
studied in detail. As a matter of fact, multimode laser systems are good
candidates to investigate noise--activated dynamics like the switching
among modes induced by quantum fluctuations (spontaneous emission)
\cite{roy}. In particular, semiconductor lasers proved to be
particularly versatile for detailed experimental  investigations of
modulation and noise-induced phenomena like stochastic resonance
\cite{gianni4,srbulk} and noise--induced phase synchronization
\cite{nips}. In those previous studies, the resonance regimes
are attained by a suitable random modulation of the bias
current which can be tuned in a well-controlled way.
It is thus natural to argue about the possibility of observing 
resonant activation with the same type of experimental setup.

In this paper, we theoretically demonstrate the phenomenon of resonant
activation in a generic rate--equations model for a two-mode semiconductor 
laser under modulation of the bias current. The basic ingredients that act in
the theoretical descriptions are a fluctuating potential barrier and some activating
noise. In the laser system, the latter is basically provided by spontaneous 
emission while current fluctuations, that appear additively into the rate
equations, effectively act multiplicatively if a suitable separation  of time
scales holds \cite{Schenzle}. In a previous paper \cite{noi}, we have explicitely
demonstrated  such multiplicative--noise effects on the mode--hopping dynamics.
This was shown by a reduction to a bistable one--dimensional
potential system with both multiplicative and additive stochastic forces.
Several predictions drawn from such a simplified model are in good agreement
with the experimental observations carried out for a bulk, Edge-Emitting Laser
(EEL) \cite{noi}. In the present context, we will show that this reduced 
description is of great help in the interpretation of simulation data.

The outline of the paper is the following. In Sec.~II we recall the
model for a two-mode semiconductor laser. In Sec.~III we present the
numerical simulation for two physically distinct cases displaying
resonant activation. These results are discussed and interpreted  by
comparing with the reduced one--dimensional Langevin model mentioned
above (Sec.~IV). We draw our conclusions in Sec.~V.

\section{Rate equations}

Our starting point is a stochastic rate-equation model for a
semiconductor laser that may operate in two longitudinal modes
whose complex amplitudes are denoted by $E_\pm$. Both of them
interact with a single carrier density $N$ that provides the necessary
amplification. The two modes have very similar linear gains, provided that
their wavelengths are almost equal and they are close to the gain
peak. Let $J(t)$ denote the bias (injection) current, the model can be
written as \cite{noi}
\begin{subequations}
\label{rateq}
\begin{equation}
\dot E_+ = \frac{1}{2}\Big[ (1 + i \alpha) g_+ - 1 \Big] E_+
+ \sqrt{2D_{sp} N}\, \xi_+ 
\label{rateq1} 
\end{equation}
\begin{equation}
\dot E_- = \frac{1}{2}\Big[ (1 + i \alpha) g_- - 1 \Big] E_-
+ \sqrt{2D_{sp} N}\, \xi_- 
\label{rateq2} 
\end{equation}
\begin{equation}
\dot N   = \gamma\Big[ J(t) - N - g_+ |E_+ |^2 - g_- |E_- |^2\Big]
\label{rateq3}
\end{equation}
\end{subequations}
where $\gamma$ is carrier density relaxation rate, $\alpha$ is 
the linewidth enhancement factor \cite{petermann}. The modal gains read
\begin{equation}
g_\pm \;=\; \frac{N \pm \varepsilon(N - N_c)}{ 1+ s |E_\pm |^2 +c
|E_\mp |^2} \; ,
\end{equation}
where $\varepsilon$ determines the difference in differential gain
among the two modes while $N_c$ defines the carrier density where
the unsaturated modal gains are equal. The parameters $s$ and $c$
are respectively the self- and cross-saturation coefficients. The
$\xi_\pm$ are two independent, complex white noise processes with
zero mean [$\langle\xi_{\pm}(t)\rangle = 0$] and unit variance
[$\langle\xi_i (t) \xi_j^*(t')\rangle = \delta_{ij} \delta(t-t')$]
that model spontaneous emission. The noise terms in Eqs.~(\ref{rateq1})
and (\ref{rateq2}) are gauged
by the spontaneous emission coefficient $D_{sp}$.

All quantities are expressed in suitable dimensionless units. In particular,
time is normalized to the photons' lifetime, which for semiconductor  laser is
typically of the order of a few picoseconds or less (see e.g.
\cite{petermann,Agrawal,sale})  

A detailed analysis of the stationary solutions of Eqs.~(\ref{rateq}) is
reported in Ref.~\cite{Albert}. 
For a constant bias current $J(t)=J_0$ and $D_{sp} = 0$, Eqs.~(\ref{rateq})
admit four different steady state solutions: the trivial one $E_{\pm} = 0$, two
single-mode solutions --- $E_+ \neq 0 , \ E_- = 0$ and viceversa --- and a
solution where both modes are lasing, $E_{\pm} \neq 0$. 
For $N_c > 1$, and $c > s$, there exist a finite interval of $J_0$ values 
for which the two single--mode
solutions coexist and are stable while the $E_{\pm} \neq 0$ is unstable
(bistable region). Here, for  $D_{sp}> 0$ the laser performs stochastic
mode-hopping, with the total emitted intensity remains almost constant while
each mode switches on and off alternately at random times.  We point out that
the emission in each mode is nonvanishing even in the ``off"  state, as the
average power spontaneously emitted in each mode at any time  is given by $4
D_{sp}N$ [recall that Eqs.~(\ref{rateq})  are usually interpreted in It\^o sense
\cite{Agrawal}]. Observation of 
this behaviour has been reported in several experimental works on  
EELs \cite{Ohtsu,Ohtsu2,IEEE}.

We remark that while Eqs.~(\ref{rateq}) aim at modeling EELs, the results
presented henceforth would apply also to polarization switching in Vertical
Cavity Surface Emitting Lasers (VCSELs). Indeed, experimental  data
\cite{gianni3} show strong similarities between this phenomenon and the
longitudinal mode dynamics. On the theoretical side,
this analogy is supported by the fact that the polarization dynamics in VCSELs
is described by models that are mathematically similar to the one discussed here
\cite{will,will2,belgi}. 

In the following, we will focus on the effect of the externally imposed 
fluctuation/modulation of the injected current. This situation
is modeled  by letting 
\begin{equation}
J(t) \;=\; J_0 + \delta J(t) \quad. 
\end{equation}
The DC value $J_0$ sets the working point and will be always 
chosen to be in the bistability region. We focus
on the case in which $\delta J$ is a Ornstein-Uhlenbeck process with
zero average $\langle\delta J(t)\rangle \;=\;0$ and correlation time
$\tau$:
\begin{equation}
\dot{ \delta J} \;=\; -{\delta J \over \tau} + \sqrt{2D_J\over \tau}
\,\xi_J
\label{ou}
\end{equation}
that means
\begin{equation}
\langle\delta J(t) \delta J(0)\rangle \;=\; {D_J} \exp(-|t|/\tau) \quad .
\end{equation}
This choice is suitable to model a finite-bandwith noise
generator. Notice that $\tau$ and the
variance of fluctuations ${D_J}=\langle\delta J^2\rangle$ can be
fixed independentely. 

Another case of experimental interest 
that we will consider is using the current modulation
\begin{equation}
\delta J \;=\; A \sin \Omega t
\label{per}
\end{equation}

To assess the nature of the stochastic process at hand, it is important to
introduce the relevant time scales.  We define first of all the switching or
relaxation time $T_R$ as the typical time for the emission to change from one mode
to the other. The main quantities we are interested in are the Kramers or residence
times $T_\pm$ defined as the average times for which the emission occurs in each
mode. In semiconductor lasers $T_\pm$ are generally much larger than $T_R$.
Typically, $T_R\sim 1-10 ns$ while residence times may range between 0.1 and
100  $\mu s$ \cite{gianni3,AppB}. The third time-scale is of course given by
the characteristic time of the external driving, namely, $\tau$ and
$2\pi/\Omega$ respectively.  

In the following, we will study how the hopping dynamics 
changes upon varying these latter parameters as well as the strength
of the perturbation.

\section{Numerical simulations}

In this Section we present the outcomes of a series of  numerical simulation of
Eqs.~(\ref{rateq}). In Ref.~\cite{noi} it was observed that the sensitivity 
of each of the $T_\pm$ on the imposed current fluctuations may be notably 
different depending on the parameters' choice. This is a typical signature 
of the multiplicative nature of the stochastic process. In particular, one can 
argue \cite{noi} that such ``simmetry-breaking" effects mostly depend on 
the ratio $\varepsilon\sigma/\delta$ where 
\begin{equation}
\sigma \;=\; {c + s \over 2}, \qquad \delta \;=\; {c - s \over 2} \quad. 
\label{delta}
\end{equation}
The parameter $\sigma$ represents the gain saturation induced by the
total power in the laser, while $\delta$ describes the reduction in
gain saturation due to partitioning of the power between the two
modes.

The possibility of obtaining qualitatively different responses depending on the
actual parameters corresponds to the different experimental observations
reported for both EELs \cite{IEEE,AppB,noi} and VCSELs \cite{gianni3,gianni4}.
Those two classes of lasers were indeed found to display markedly different
simmetry-breaking  effects under current modulation. To account for those
features, we consider two different sets of phenomenological parameters.  For
definiteness, in both cases we fix $\varepsilon=0.1$, $s=1.0$, $N_c=1.1$,
$\gamma=0.01$ and change the values of $c$ and $D_{sp}$ (see Table I). The first
set ($\delta=0.05$) corresponds to the case in which added modulation changes
the hopping time scale in an almost symmetric way. On the contrary, in the
second case ($\delta=0.15$) the asymmetry effect of the noise is stronger
\cite{noi}. We can thus consider the two as representative of the VCSELs and
EELs case respectively. The value of $J_0$ has been empirically adjusted to
yield $T_+\simeq T_- \equiv T_s$ and an almost symmetric distribution of
intensities in absence of modulation. The actual values are about 10\% above the
laser threshold.  The spontaneous emission coefficient $D_{sp}$ has been chosen
to yield a value of the residence times of the same order of magnitude of the 
experimental ones. 

In the following, we decide to set $\alpha=0$ which is appropriate
for our EEL model where the phase dynamics is not relevant \cite{noi}.
This choice may however not be fully justified for the VCSEL case. 
In this respect, the simulations presented below are representative of 
the VCSEL dynamics only in a qualitative sense. Nonetheless, it 
should  be pointed out that a 1D Langevin model independent of $\alpha$ 
describes also the VCSEL case ~\cite{will,will2}.
Since resonant activation is mainly due to the multiplicative
noise effect described by such equations [see Eq.~(\ref{phidot}) below] 
we consider this as an indirect proof that phenomenology we will report
below should be observable also in the VCSEL case.

The largest part of the simulations were performed
with Euler method with time steps 0.01-0.05 for times in the 
range $10^7 - 10^8$ time units depending on the 
values of $\tau$ and $\Omega$. For comparison, some
checks with Heun method \cite{toral} have also been carried on.
Within the statistical accuracy, the results are found to be
insensitive the the choice of the algorithm.

\begin{table}
\caption{
\label{tab:table1}
The parameter values used in the two series of 
simulations of Eqs.~(\ref{rateq}), the other values are given 
in the text. }
\begin{ruledtabular}
\begin{tabular}{ccccc}
$c$ & $D_{sp}$ & $J_0$ & $\delta$ & $\sigma$ \\
\hline
1.1 & $0.7 \times 10^{-5}$ & 1.197 & 0.05 & 1.05 \\
1.3 & $1.5 \times 10^{-5}$ & 1.194 & 0.15 & 1.15 \\
\end{tabular}
\end{ruledtabular}
\end{table}

\subsection{Stochastic modulation}

\begin{figure}[h]
\begin{center}
\includegraphics[clip,width=8cm]{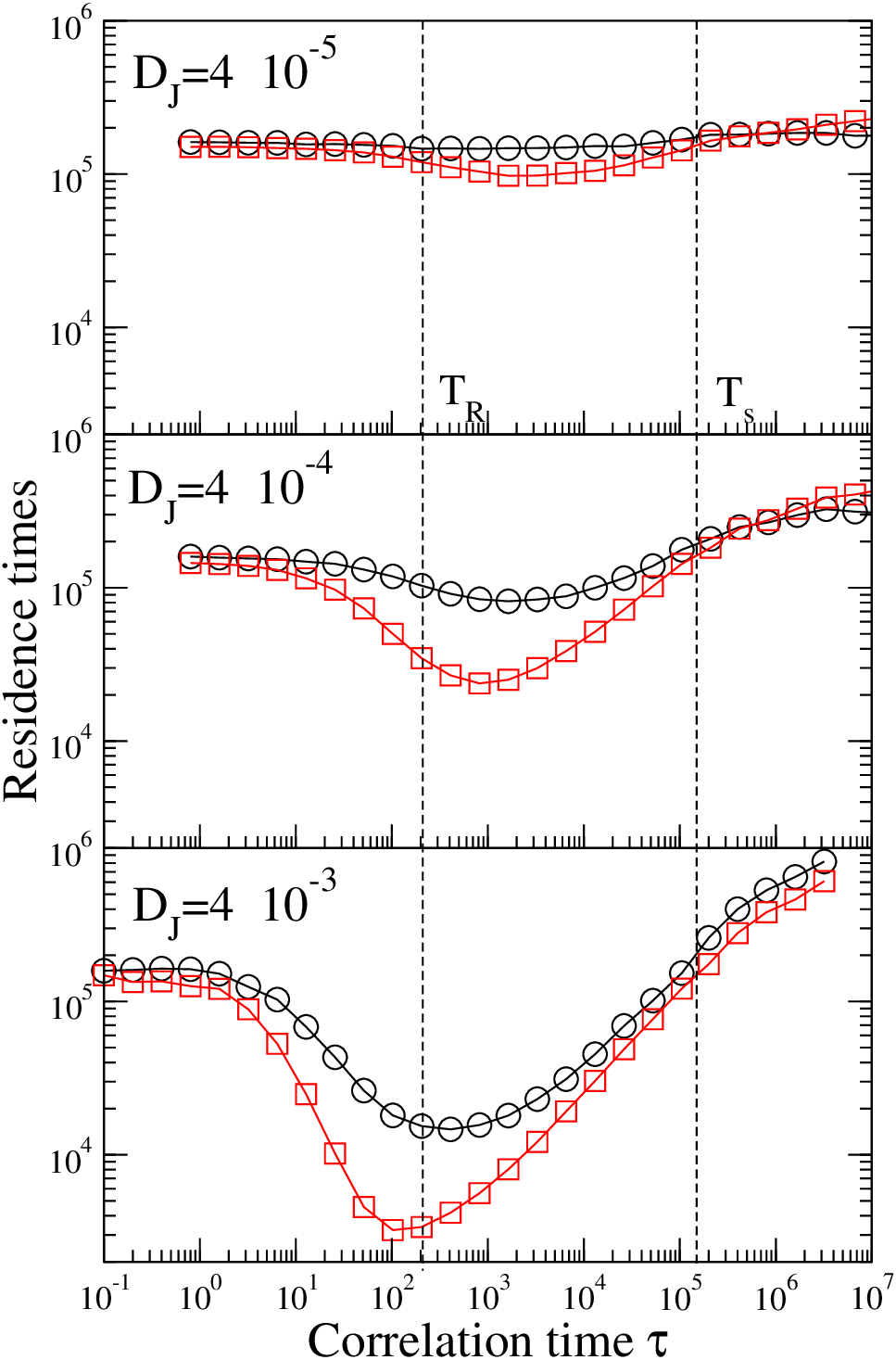}
\caption{
(Color online)
Simulations of the rate equations with Ornstein-Uhlenbeck 
current fluctuations, parameter set with $c=1.1$ (see text
and Table~I): residence times $T_+$ (squares) and $T_-$ (circles) 
for increasing values of the current variance $D_J$.
The values of the relaxation time $T_R$ and
the hopping time $T_s$ (in absence of modulation) are marked 
by the vertical dashed lines.
}
\label{f:stoc1}
\end{center}
\end{figure}

\begin{figure}[h]
\begin{center}
\includegraphics[clip,width=8cm]{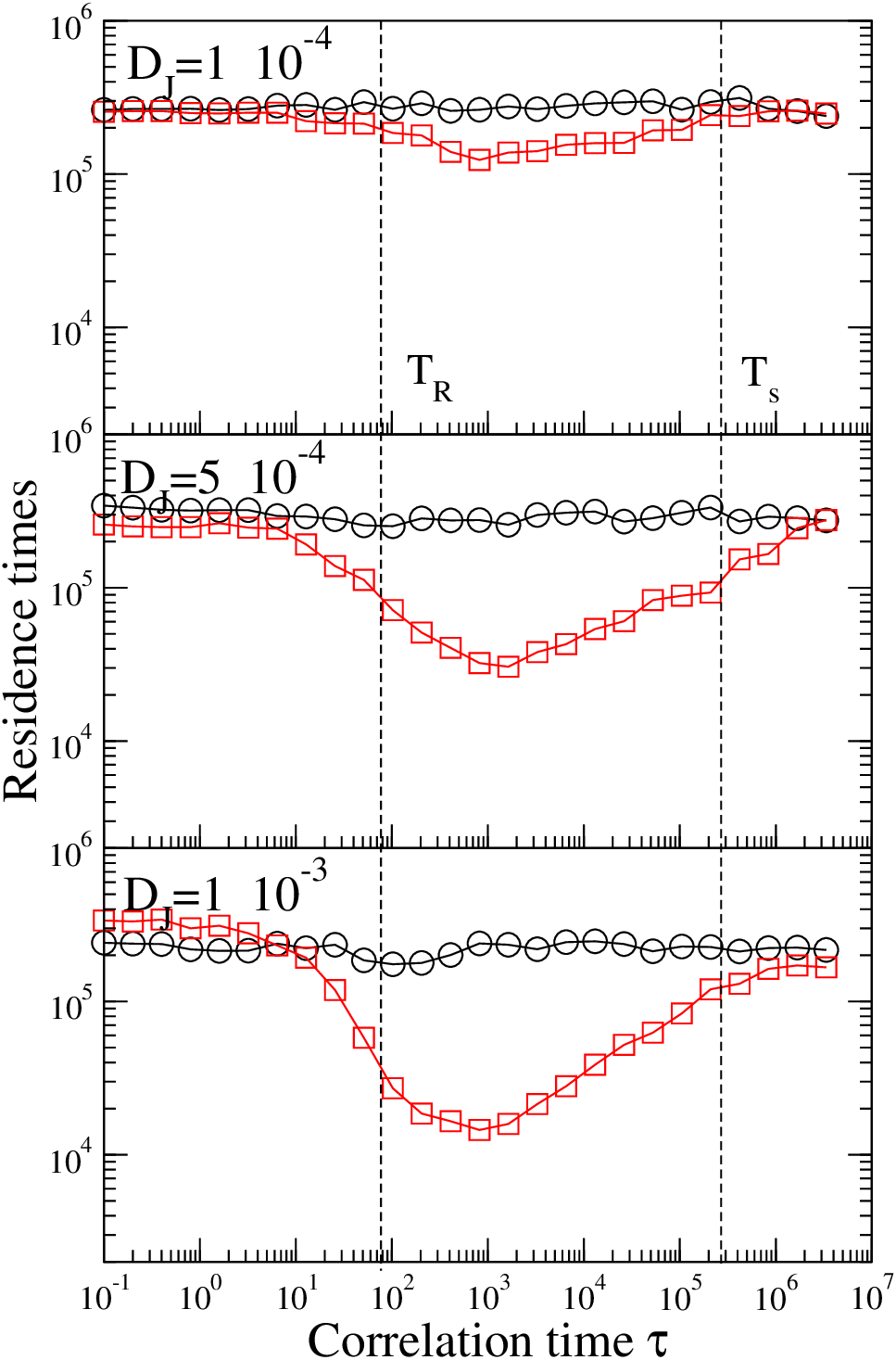}
\caption{
(Color online)
Simulations of the rate equations with Ornstein-Uhlenbeck 
current fluctuations, parameter set with $c=1.3$ (see text
and Table~I): residence times $T_+$ (squares) and $T_-$ (circles) 
for increasing values of the current variance $D_J$.
}
\label{f:stoc2}
\end{center}
\end{figure}

Let us start illustrating the results in the case of stochastic  current
modulation (Eq.~(\ref{ou})). In Fig.~\ref{f:stoc1} and \ref{f:stoc2} we report
the measured dependence  of the residence times $T_\pm$ on the correlation  time
$\tau$ for the two parameter sets given in  Table I and different values of the
noise variance $D_J$. In all cases, the curves display well-pronounced minima at
an optimal value of $\tau$.  This is the typical signature of resonant
activation. The minima are almost located between the relaxation time $T_R$ and
the hopping time  $T_s$ (marked by the vertical dashed lines). The values of
$T_R$ reported in the figures have been estimated from the reduced model
discussed  in the next Section, see Eq.~(\ref{tr}) below.

The effect manifest in a different way for the second parameter set. In the case
of Fig.~\ref{f:stoc1} \textit{both} times attain a minimum, albeit with
different values. On the contrary the data of Fig.~\ref{f:stoc2} show that one
of the  two times is hardly affected from the external perturbation regardless
of the value of $\tau$. In other terms, we can tune the current correlation in
such a way that  emission along only one of the two modes is strongly  reduced
(about a factor 10 in the simulation discussed  here). 

\subsection{Periodic modulation}

\begin{figure}[h]
\begin{center}
\includegraphics[clip,width=8cm]{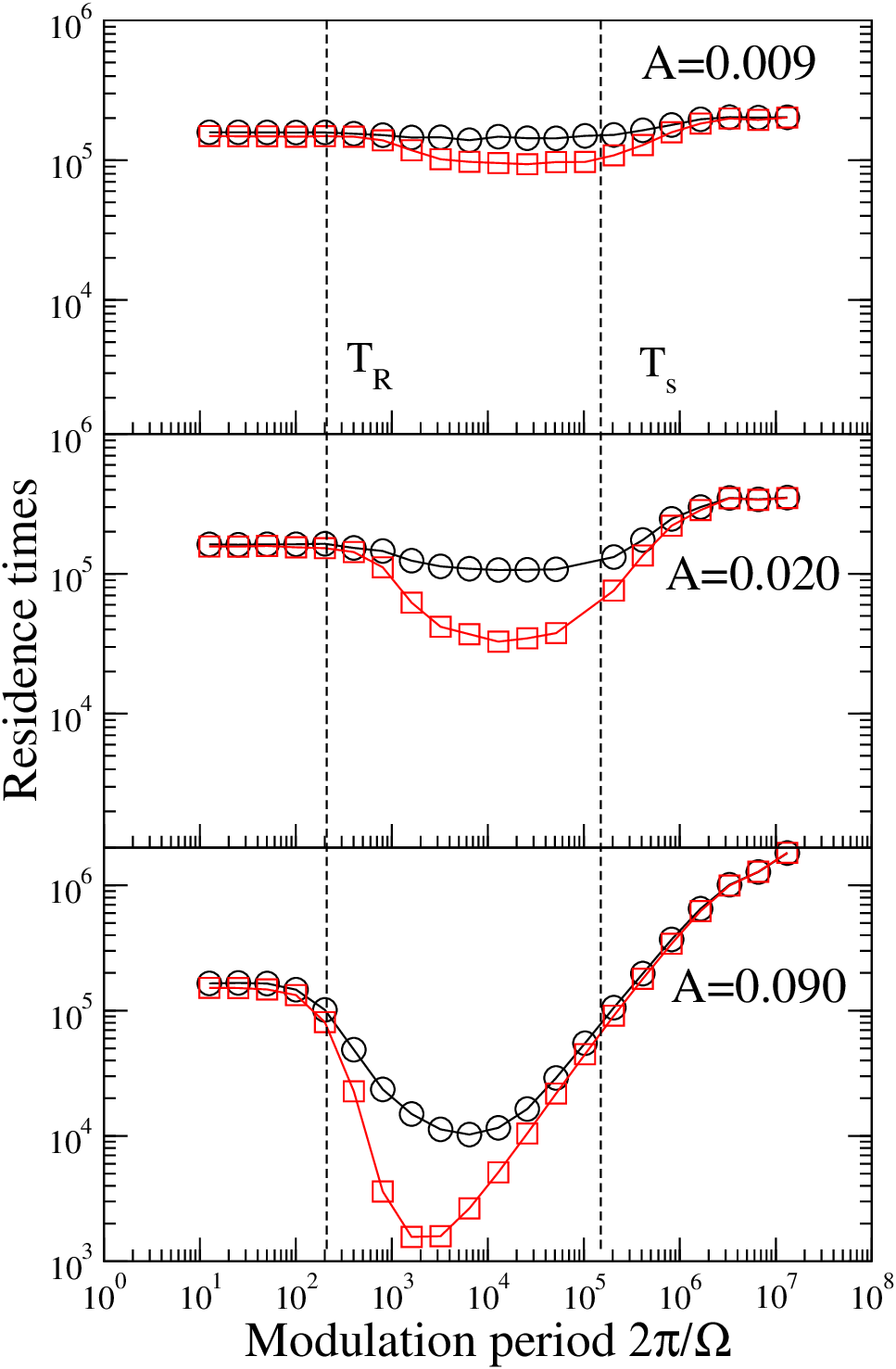}
\caption{
(Color online)
Simulations of the rate equations with sinusoidal  
modulation of the current, parameter set with $c=1.1$ (see text
and Table~I): residence times $T_+$ (squares) and $T_-$ (circles) 
for increasing values of modulation amplitude $A$.
}
\label{f:per1}
\end{center}
\end{figure}

\begin{figure}[h]
\begin{center}
\includegraphics[clip,width=8cm]{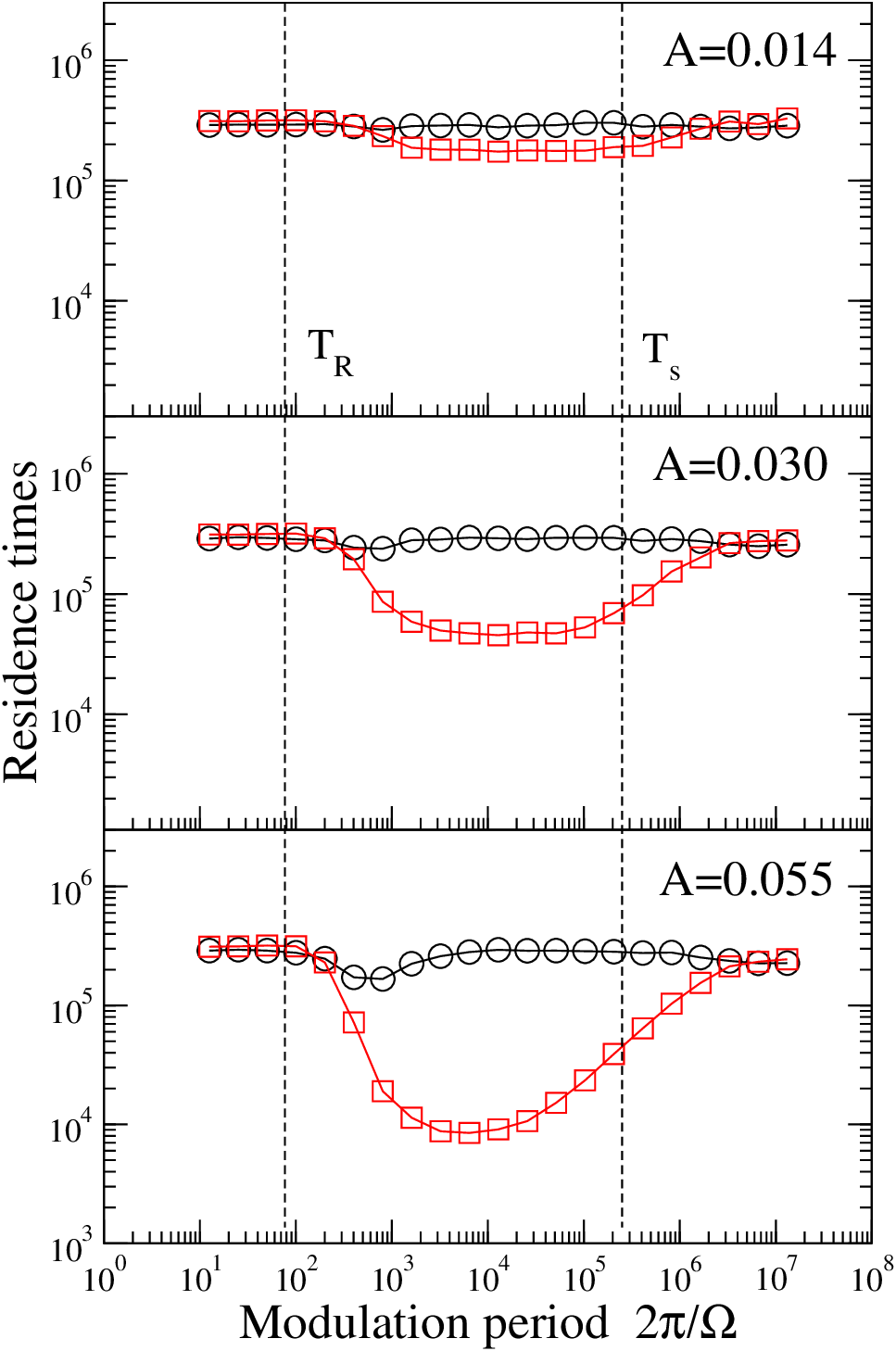}
\caption{
(Color online)
Simulations of the rate equations with sinusoidal  
modulation of the current, parameter set with $c=1.3$ (see text
and Table~I): residence times $T_+$ (squares) and $T_-$ (circles) 
for increasing values of modulation amplitude $A$.
}
\label{f:per2}
\end{center}
\end{figure}

Let us now turn to the case of sinusoidal current modulation (Eq.~\ref{per}).
In Fig.~\ref{f:per1} and \ref{f:per2} we report the measured dependence  of the
residence times $T_\pm$ on the frequency  $\Omega$ for the two parameter sets
given in  Table I and different values of the amplitude $A$. For comparison with
the previous case we choose $A$ such that the RMS value of (\ref{per}) is
roughly equal to the variance of (\ref{ou}),  i.e. $A\simeq \sqrt{2D_J}$.

As in the previous case, the curves display resonant
activation at an optimal value of $\Omega$. For the second set of parameters,
one of the two hopping times is more reduced than the other
(compare Fig.~\ref{f:per2} with Fig.~\ref{f:stoc2}).
It should be also noticed that the data in Fig.~\ref{f:stoc2} display
some statistical fluctuations while the curves for the periodic 
modulation are smoother.

\section{Insights from a reduced model}

In order to better understand the activation phenomenon it
is useful to reduce the five--dimensional dynamical system
(\ref{rateq}) to an effective one-dimensional system. 
This has been accomplished in Ref.~\cite{noi}. For completeness, 
we only recall here some basic steps of the derivation. 
In the first place, we introduce the change of coordinates 
\begin{equation}
E_+ = r \cos \phi \exp i\psi_+,\qquad  
E_- = r \sin \phi\exp i\psi_- \; .
\end{equation}
In these new variables, $r^2$ is the total power emitted by
the laser, and $\phi$ determines how this power is partitioned
among the two modes. The values $\phi=0,\pi/2$ correspond to
pure emission in mode $+$ and $-$ respectively. The phases 
$\psi_\pm$ do not influence the evolution of the modal
amplitudes and carrier density and can be ignored. 

In order to simplify the analysis, we assume that (i) The difference
between modal gains is very small, i.e., $N_c \gtrsim 1$, $\varepsilon
\ll 1$,  $c\gtrsim s$; (ii) the laser operates close enough to threshold, so
that $r^2 \ll 1$ and the saturation term is small: in this limit, 
$r$ and $N$ decouple to leading order from $\phi$;
(iii) $r$ and $N$ can be adiabatically eliminated and (iv) only their
fluctuations around the equilibrium values due to $J$ are retained.
This last assumption holds for weak spontaneous noise and amounts
to say that $r$ and $N$ are stochastic processes given by nonlinear
transformations of $J$ (see Eqs.~(16) in Ref.~\cite{noi}). This requires
that $J$ does not change too fast. For example, in the case of the
Orstein--Uhlenbeck process, Eq.~(\ref{ou}), $\tau$ should be  larger than
the relaxation time of the total intensity. The validity  of the above
reduction has been carefully checked against simulations of the
complete model \cite{noi}. For the scope of the present work, we
performed a further check  by comparing the spectrum of fluctuations
of $r^2$ with the imposed one, Eq.~(\ref{ou}). Indeed, the behaviour is
the same for $\tau>T_R$ while for shorter $\tau$ some differences are
detected. This means that the reduced description discussed below
becomes less and less accurate. On the other hand, in this regime
spontaneous fluctuation should dominate and this limitation become 
less relevant for our purposes.

Altogether, the hopping dynamics is effectively one-dimensional and 
is described by the slow variable $\phi$. Its evolution is ruled 
by the effective Langevin equation 
\begin{equation}
\dot \phi\; =\;
-\frac{1}{2}\Big[a \cos2\phi + b\Big]\sin2\phi \,+\, {2D_\phi\over \tan 2\phi}
 + \sqrt{2 D_\phi}\, \xi_\phi
\label{phidot}
\end{equation}
where, together with (\ref{delta}) we have defined the new set of parameters
\begin{eqnarray}
&& J_s \;=\; {(1+\sigma) N_c -1 \over \sigma}\\
&& a \;=\; {\delta  \over 1+\sigma}\,(J-1) \\
&& b\;=\; {\varepsilon \sigma \over 1+\sigma}\,(J-J_s) \\
&& D_\phi \;=\;\frac {( 1 + \sigma J)^2} {(1 + \sigma) (J-1)}
   \, D_{sp}\quad.
\label{params}
\end{eqnarray}

We remind in passing that the same equation (\ref{phidot}) has been
derived by Willemsen et al.~\cite{will,will2} to describe polarization
switches in VCSELs (see also Ref.~\cite{VanDerSande} for a similar 
reduction). The starting point of their
derivation is the San Miguel-Feng-Moloney model~\cite{Feng}. The
physical meaning of the variable $\phi$ is different from here as it
represents the polarization angle of emitted light. This supports the
above claim that, upon a suitable reinterpretation of variables and
parameters, many of the results presented henceforth may apply also to
the dynamics of VCSELs.

In absence of modulation ($\delta J=0$), Eq.~(\ref{phidot})
is bistable in an interval of current values where it admits
two stable stationary solutions $\phi_\pm$ and an unstable one 
$\phi_0$ (double-well). This regime correspond 
to the bistability region of model (\ref{rateq}). 
Notice that for $J_0=J_s$, $b=0$ the hopping between the two 
modes occurs at the same rate. The above definitions allows 
an estimate of relaxation time $T_R$ defined above. This is 
is the inverse of the curvature of the potential in $\phi_0$. 
For $J_0=J_s$ this is straightforwardly evaluated to be
\begin{equation}
T_R \; \simeq \; \frac{(1+\sigma)}{\delta(J_s - 1)}
\label{tr}
\end{equation}
For the two parameter sets given in Table I 
one finds $T_R = 210$ $T_R = 77.0$, respectively. These are 
the values emploied to draw the leftmost vertical lines  
in Figs.~\ref{f:stoc1}-\ref{f:per2}.

The effect of a time-dependent current is to make the coefficients 
$a$, $b$ and $D_\phi$ fluctuating. It can be shown \cite{noi} 
that the effect on  $D_\phi$ can be recasted as a renormalization of 
the intensity of the spontaneous-emission noise. However, for the 
parameters employed in the present work it turns out that this correction 
is pretty small and will be neglected henceforth by simply considering 
$D_{\phi}$ as constant \cite{note}. For simplicity, we also 
disregard the dependence of $D_\phi$ on $\delta J$ in the drift 
term of Eq.~(\ref{phidot}). Under those further simplifications 
the Langevin equation can be rewritten as
\begin{equation}
\dot \phi \;=\; - U'(\phi) - V'(\phi)\,\delta J
+ \sqrt{2 D_\phi} \, \xi_\phi
\label{langmu}
\end{equation}
where we have express the force term as derivatives of 
the ``potentials"
\begin{eqnarray}
 U(\phi) &\;=\;& -{\delta (J_0-1)\over 16(1+\sigma)}\cos 4\phi 
- {\varepsilon \sigma (J_0-J_s)\over 4(1+\sigma)}\cos 2\phi 
\nonumber \\
&& - D_\phi \ln \sin 2\phi \\
\label{uphi}
 V(\phi) &\;=\;& -{\delta\over 16(1+\sigma)} \cos 4\phi -
{\varepsilon\sigma\over 4(1+\sigma)}\cos 2\phi .
\end{eqnarray}
Langevin equations of the form (\ref{langmu}) with (\ref{ou})
have been thoroughly studied in the literature (see e.g.
\cite{h94,h95,r95,iw96,m96} and references therein) as prototypical
examples of the phenomenon of activated escape over a fluctuating
barrier. In view of their non-Markovian nature, their
full analytical  solution for arbitrary $\tau$ is not generally
feasible. Several approximate results can be provided in some
limits. 

For an arbitrary choice of the parameters, $V$ has a different
symmetry with respect to $U$ meaning that the effective amplitude of
multiplicative noise is different within the two potential wells. If
this difference is large enough, current fluctuation will remove the
degeneracy between the two stationary solutions.
This is best seen by computing the istantaneous potential barriers 
$\Delta U_\pm (t)$ close to the 
symmetry point $J_0 = J_s$ . For weak noise and $\delta J \ll (J_s-1)$,  
they are given to first-order in $\delta J(t)$ by
\begin{equation}
\Delta U_\pm (t) \;\simeq\; \frac{\delta}{8(1+\sigma)}(J_s-1)
\,+\, \frac{\delta \pm 2\varepsilon\sigma}{8(1+\sigma)}
\delta J(t) \, .
\label{fbar}
\end{equation}
Obviously, this last expression makes sense only when the
fluctuating term is sub-threshold i.e. whenever the system is
bistable. In the case of periodic modulation, formula (\ref{fbar}) 
allows estimating the range of amplitude values for a sub-threshold 
driving
\begin{equation}
A \;<\; \frac{\delta (J_s-1)}{\delta \pm 2\varepsilon\sigma}\quad.
\end{equation}
Using this condition, along with the parameter values at hand, we 
deduce that the cases displayed in lower panels of 
Figs.~\ref{f:per1} and \ref{f:per2} correspond to superthreshold driving.
However, while the minima are much more pronounced than in the 
other panels, there is no qualitative difference in the system response.
In the case of stochastic modulation, the same remark applies in 
a probabilistic sense for the last panels of Figs.~\ref{f:stoc1} and 
\ref{f:stoc2}.

Altogether, the mode switching can be seen as an activated escape
over fluctuating barriers given by Eq.~(\ref{fbar}). The statistical
properties of the latter process is controlled by the current
fluctuations. We now discuss the properties of various regimes.
For simplicity, we refer to the case of stochastic modulations.
Most of the remarks and formulas reported in the following 
Subsection should apply also to the periodic case by 
replacing $\tau$ and $D_J$ with $2\pi/\Omega$ and $A^2/2$ whenever 
appropriate.
 
\subsection{Fast barrier fluctuations: $ \tau < T_R \ll  T_\pm$}

As we already pointed out, in this regime the reduction to
Eq.~(\ref{langmu}) is not justified. We may thus only expect some qualitative
insight on the behaviour of the rate-equations. From a mathematical point of 
view, some analytical approximations for equations like (\ref{langmu}) are
feasible in this limit (see e.g. Ref.~\cite{h95} for the stochastic case). For
our purposes, it is sufficient to note that in this regime the effect of $\delta
J$ is hardly detected  for both types of driving (see again
Figs.~\ref{f:stoc1}-\ref{f:per2}). Note also that working at $D_J$ fixed means
that for $\tau \to 0$ the  fluctuation become negligible.

\subsection{Resonant activation: $T_R < \tau \ll T_\pm$}

If $T_R < \tau$ we are in the colored noise case. The problem is amenable of
a kinetic description which amounts to neglect intrawell motion and reduce
to a rate model describing the statistical transitions in terms of
transition rates. If we consider $\tau$ as a time scale of the external
driving we can follow the terminology of Ref.~\cite{talk} and refer to this
situation as the ``semiadiabatic" limit of Eq.~(\ref{langmu}). 

In this regime, the residence time is
basically the shortest escape time, which in turn correspond to the
lowest value of the barrier (the noise is approximatively constant
in the current range considered henceforth). For the case of
interest,  $\delta < 2\varepsilon \sigma$ we can use (\ref{fbar}) to
infer that the minimal values of $\Delta U_\pm$ should be attained
for $\delta J \propto \mp\sqrt{D_J}$ respectively. This yields
\begin{equation}
T_\pm \;\simeq\;
T_s \exp\Big[-K \frac{2\varepsilon\sigma\pm\delta}{1+\sigma}
\,\frac{\sqrt{D_J}}{D_\phi} \Big]
\label{hopt}
\end{equation}
where $K$ is a suitable numerical constant. 
Notice that $\delta$ controls the asymmetry level: if $\delta \ll
2\varepsilon\sigma$ the two residence times decrease at
approximatively the same rate. This prediction is verified in the
simulations and also in the experiment \cite{noi}.

\begin{figure}[h]
\begin{center}
\includegraphics[clip,width=8cm]{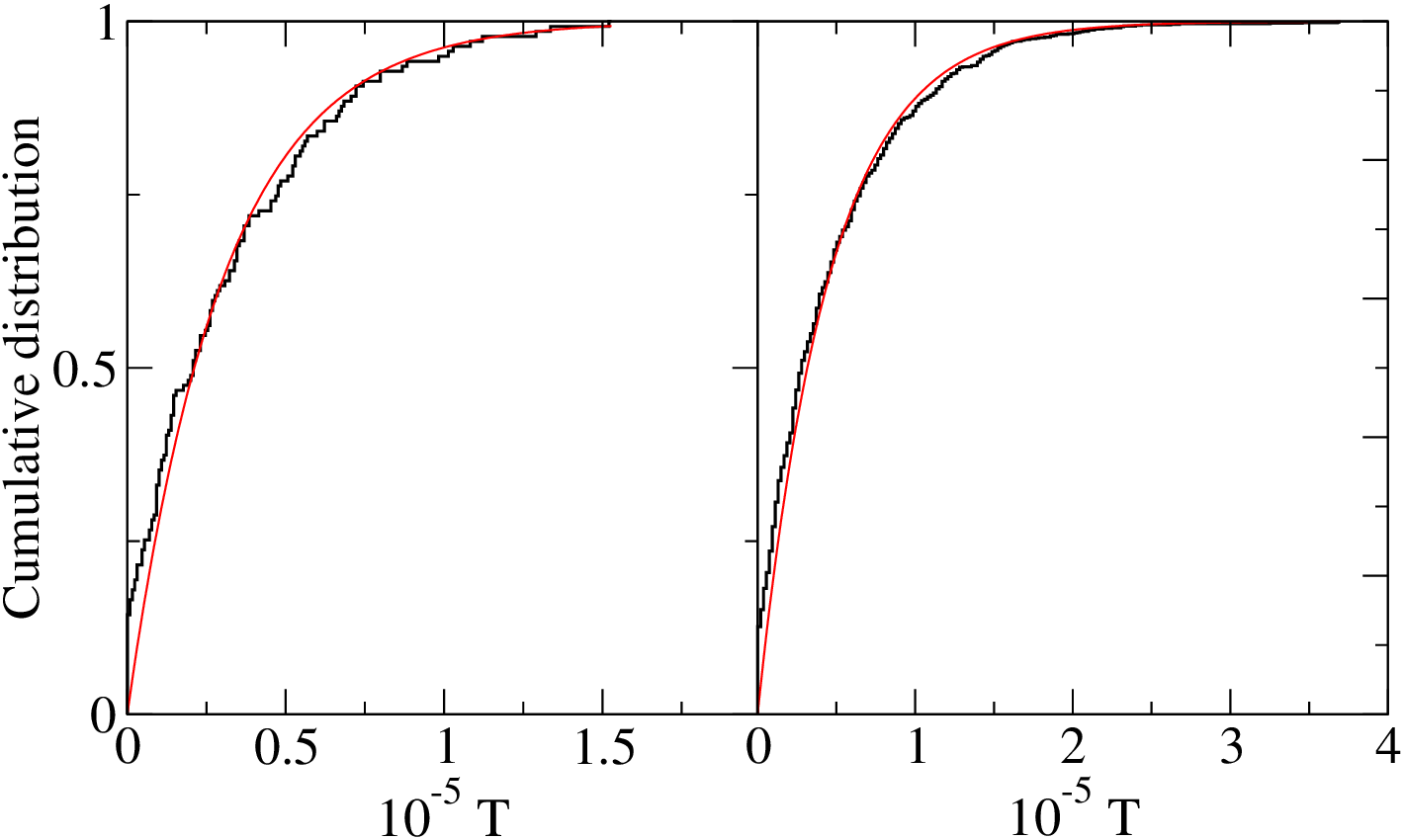}
\caption{
(Color online)
Cumulative distributions of the residence times in the 
resonant activation region, parameter set with $c=1.3$
(see text and Table~I). Left panel: stochastic modulation
with $D_J=5 \times 10^{-4}$, $\tau=1.638 \times 10^3 $. 
Right panel: periodic modulation  with $A=0.03$ and 
period $1.286 \times 10^4 $. We report only the histograms for the 
times whose averages are denoted by $T_+$ in the text.
Solid line is the cumulative Poissonian distribution 
with the same average.
}
\label{f:cum}
\end{center}
\end{figure}

As a further argument in support of the above reasoning, we also 
evaluated the probability distributions of the residence times 
obtained from the simulation of the rate equations. In Fig.~\ref{f:cum}, 
we show two representative cumulative distributions. The data are 
well fitted by a Poissonian $P(T)=1-\exp(-T/T_\pm)$ for both the 
stochastic and periodic modulation cases. This confirms that hopping 
occours preferentially when a given (minimal) barrier occurs.  

\subsection{Slow barrier, frequent hops: $T_R \ll T_\pm \ll \tau$}

This corresponds to the adiabatic limit in which the time scale of the 
external driving is slower than the intrinsic dynamics of the 
system \cite{talk}. To a first approximation we can here treat current 
fluctuations in a parametric way. 
Correction terms may be evaluated by means of a suitable perturbation
expansion in the small parameter $1/\tau$ \cite{iw96}. 
If $\delta J$ is small enough  for the   
expression (\ref{fbar}) to make sense, the escape time can be 
estimated as  
the average of escape times over the distribution of barrier 
fluctuations, i.e. $\langle T_\pm \rangle_{\delta J}$. 
For the case of Eq.~(\ref{ou}), the variable $\delta J$ is Gaussian 
and we can use the identity
$\langle \exp{\beta z}\rangle= \exp(\beta^2\langle z^2\rangle /2)$
to obtain \cite{m96}
\begin{equation}
T_\pm \;\simeq\; T_s \exp\Big[
{2(\delta\pm 2\varepsilon\sigma)^2 \over (1+\sigma)^2 D_\phi^2} \,
D_J \Big] .
\label{largetau}
\end{equation}
This reasoning implies that for large $\tau$ the residence times 
should approach two different constant values. 
A closer inspection of the graphs (in linear scale) reveals that
this is not fully compatible with the data of Fig.\ref{f:stoc1} even 
for the smallest value of $D_J$. In several cases, $T_\pm$ continue to 
increase with $\tau$ and no convincing evidence 
of saturation is observed. We note that the same type of 
behaviour was already observed in the analog simulations data 
of Ref.~\cite{m96}. There, an increase of hopping times duration 
at large $\tau$ was found. The Authors of Ref.~\cite{m96} explained
this as an effect of a too large value of the noise fluctuation
forcing the system to jump roughly every $\tau$. We argue that 
the same explanation holds for our case. This is also consistent
with the fact that the exponential factors in Eq.~(\ref{largetau}) 
evaluated with the simulation parameters turn out to be much larger 
than unity.

\section{Conclusions}

In this paper, we have explored numerically and analytically the effects of
external current fluctuations on the mode-hopping dynamics  in a model of a
bistable semiconductor laser.  To the best of our knowledge, this setup
provides the first theoretical evidence of resonant activation in a laser
system. As the phenomenon has hardly received any experimental confirmation  in
optics, we believe that our study may open the way to future  research in this
subfield.

The model we investigated is based on a rate-equation
description, where the bias current enters parametrically into the evolution of
the modal amplitudes. We considered, two kinds of current flutuations, namely, a
stochastic process ruled by an Orstein-Uhlenbeck statistics, and a coherent,
sinusoidal modulation. These choices are motivated by the aim of proposing a
suitable setup for an experimental verification of our results. Upon varying the
characteristic time-scale of the imposed fluctuations, we have shown that the
residence times attain a minimum for a well-defined value, which is the typical
signature of resonant activation. The magnitude of the effect can be different
depending on the parameters of the model. Moreover, the response of the
system appears very much similar for both periodic and random modulations. 

The reduction of the rate equations to a one-dimensional Langevin equation
allowed  us to recast the problem as an activated escape over a fluctuating
barrier.  To first approximation, the fluctuating barrier (multiplicative term)
is mainly controlled by current modulations while the spontaneous noise act as
an additive source. This simplified description has allowed us to draw some
predictions (e.g. the dependence of residence times on noise strength) and to
better understand the role of the physical parameters.  Given the generality of
the description, our results should apply to a broad class of multimode lasers,
including both Edge Emitting and Vertical Cavity Lasers.

From an experimental point of view, driving the laser in a orders-of-magnitude
wide range of time-scales is more feasible in the case of a sinusoidal
modulation than for a colored, high frequency noise. However, given the evidence
of a resonant activation phenomenon for such modulation, our results indicate
that it occurs almost for the same parameters in the case of colored noise,
provided that the RMS of the modulations equals the amplitude of the added
noise. Thus, the phenomenon could be fully exploited along those lines.   
Since the reported experimental evidences of the phenomenon are so far 
scarce, we hope that the present work could suggest a detailed characterization
in optical systems that allows for both very precise measurements and 
careful control of parameters.


\begin{thebibliography}{99}
\bibitem{note} For specific choices of the parameters, 
this approximation may not be justified. For example, 
when $\delta \simeq 2\varepsilon\sigma$ the barrier fluctuations
$\Delta U_-$ is hardly affected by a change in $J$ and the 
renormalization of spontaneous noise cannot be neglected. 
Since the parameters are independent we restrict to the 
generic case in which the above condition is not fulfilled.
\end{thebibliography}
\end{document}